\documentclass[9pt,twoside]{rmaa-rho}
\RMxAAtemplatetype{\RMxAA} 

\usepackage[justification=justified,singlelinecheck=false]{caption}

\vol{ }
\pages{ }
\thisyear{ }
\doi{}{}

\title{The Phenomenological Nature of Quasar-type Blazars (BZQ). I. Revisiting the Flat-Spectrum Paradigm}

\author[1]{Jonhatan U. Guerrero-González \orcidlink{0009-0002-2044-3274}}
\author[1]{Vahram Chavushyan \orcidlink{0000-0002-2558-0967}}
\author[1,2]{Victor M. Patiño-Álvarez \orcidlink{0000-0002-5442-818X}}

\affil[1]{Instituto Nacional de Astrofísica, Óptica y Electrónica, Luis Enrique Erro \#1, Tonantzintla, Puebla, México, C.P. 72840.}
\affil[2]{Max-Planck-Institut für Radioastronomie, Auf dem Hügel 69, D-53121 Bonn, Germany.}

\leadauthor{Guerrero-González et al.}
\smalltitle{The Phenomenological Nature of Quasar-type Blazars (BZQ). I.}

\corres{Jonhatan U. Guerrero-González}
\email{jonhugg13@gmail.com}

\received{September 26th, 2025}
\accepted{May 11th, 2026}

\license{Texto de la licencia aquí}

\setbool{rho-abstract}{true} 
\setbool{rho-resumen}{true} 

\begin{abstract}
    We reevaluate 610 sources classified as Flat-spectrum radio quasars (FSRQs) in the 5th edition of the Roma-BZCAT. Optical spectra from SDSS DR16 confirm broad emission lines within $0.11 \leq z \leq 5.28$. To assess their blazar-like behavior, we combine ZTF optical variability with radio morphologies from FIRST, LOFAR, and VLBI, defining Confirmed, Possible, and Non-Confirmed BZQs. Rest-frame 1.4--10 GHz radio spectra were homogenized and fitted with error-weighted power laws. We show that the scheme of \citet{Park2013} often misclassifies nearly flat spectra as inverted and some prominently steep spectra as flat. Using the individual uncertainty $\sigma_{\alpha,i}$, we classify spectra as flat if $|\alpha_i| \leq 2\sigma_{\alpha,i}$, prominently steep if $\alpha_i > 2\sigma_{\alpha,i}$, and inverted if $\alpha_i < -2\sigma_{\alpha,i}$. This criterion, defined source by source and intended as a phenomenological classification for this sample, better reflects the observed spectral shapes and confirms that most BZQs are consistent with being flat within measurement precision, although a non-negligible fraction departs from strict flatness. We also classify full spectral morphologies as power-law, peaked, restarted-peaked, or inverted-peaked, associated with distinct jet processes and activity cycles. About 60\% of the sources with at least two decades of frequency coverage exhibit restarted-peaked spectra, suggesting recurrent jet activity. The observed diversity indicates that the label ``\textit{flat-spectrum radio quasar}'' does not fully describe this population, and that the more general term BZQ may better reflect its phenomenological diversity.
\end{abstract}


\begin{resumen}
    Reevaluamos 610 fuentes clasificadas como cuásares de espectro plano en radio (FSRQs) en la 5ta edición del Roma-BZCAT. Los espectros ópticos de SDSS DR16 confirman líneas de emisión anchas en $0.11 \leq z \leq 5.28$. Para evaluar su comportamiento tipo blazar, combinamos variabilidad óptica de ZTF con morfologías de radio de FIRST, LOFAR y VLBI, definiendo BZQ Confirmados, Posibles y No Confirmados. Los espectros de radio en el marco de reposo 1.4--10 GHz fueron homogeneizados y ajustados mediante leyes de potencia ponderadas por errores. Mostramos que el esquema de \citet{Park2013} suele clasificar erróneamente como invertidos a espectros casi planos y como planos a algunos espectros con inclinación prominente. Usando la incertidumbre individual $\sigma_{\alpha,i}$, clasificamos los espectros como planos si $|\alpha_i| \leq 2\sigma_{\alpha,i}$, con inclinación prominente si $\alpha_i > 2\sigma_{\alpha,i}$, e invertidos si $\alpha_i < -2\sigma_{\alpha,i}$. Este criterio, definido fuente por fuente y propuesto como una clasificación fenomenológica para esta muestra, refleja mejor las formas espectrales observadas y confirma que la mayoría de los BZQ son planos dentro de la precisión de medición, aunque una fracción no despreciable se aparta de la planitud estricta. También clasificamos morfologías de espectro completo como Ley de potencias, con pico, pico reactivado o pico invertido, asociadas a distintos procesos de chorro y ciclos de actividad. Alrededor del 60\% de las fuentes con al menos dos décadas de cobertura en frecuencia muestran espectros de pico reactivado, sugiriendo actividad recurrente del chorro. La diversidad observada indica que ``\textit{cuásar de espectro plano en radio}'' no describe completamente esta población y que el término más general BZQ puede reflejar mejor esta diversidad fenomenológica.
\end{resumen}

\begin{document}

\nolinenumbers

\maketitle
\pagestyle{fancy}\thispagestyle{firststyle}


\section{INTRODUCTION}
\label{sec:Introduction}

    \RMxAAstart{A}ctive galactic nuclei (AGNs) are among the most luminous sources in the Universe, powered by accretion of matter onto supermassive black holes. A fraction of AGNs are classified as radio-loud owing to the presence of relativistic jets, whose emission can dominate across the entire electromagnetic spectrum \citep{Urry1995, Padovani2017}. When such jets are oriented close to our line of sight, the object is observed as a blazar \citep{Blandford1978, Urry1995}. Relativistic beaming and Doppler boosting make blazars appear exceptionally luminous and highly variable at all wavelengths, from radio to $\gamma$-rays \citep{Ulrich1997, Aharonian2007, Hovatta2019}.

Blazars are traditionally divided into two main subclasses: BL Lacertae objects (BL Lacs) and Flat-spectrum radio quasars. Early classification schemes relied on the equivalent width of optical emission lines, adopting EW $<$ 5 \AA\ as the dividing criterion \citep{Stickel1991, Stocke1991}. In this work, we follow the nomenclature of the Roma-BZCAT catalog \citep{Massaro2009}: BZB for BL Lacs, i.e. AGNs with nearly featureless optical spectra or spectra dominated by host-galaxy absorption lines and weak, narrow emission lines; and BZQ for FSRQs, characterized by broad emission lines and prominent blazar signatures. Typically, BZQs exhibit synchrotron peaks in the infrared, while BZBs span a wider range, from the infrared to the X-ray band \citep{Abdo2010a, Hovatta2019}. Increasing evidence, however, indicates that BZBs and BZQs represent a continuum of physical properties, such as accretion rate, black hole mass, and jet power, rather than two sharply distinct populations \citep{Ghisellini2011, Padovani2017}.

The spectral energy distribution (SED) of blazars is characterized by two broad components \citep{Fossati1998, Abdo2010b}. The low-energy bump, extending from radio to optical/UV or X-rays, arises mainly from synchrotron radiation emitted by relativistic particles in the jet. The high-energy bump, spanning from X-rays to $\gamma$-rays, is most commonly attributed to inverse Compton scattering in leptonic models \citep{Maraschi1992, Sikora1994, Boettcher2013}. Competing hadronic models, invoking proton synchrotron and photo-hadronic interactions, have also been proposed \citep{Mucke2003, Boettcher2019}. In either scenario, the contribution of internal and external photon fields, including the accretion disk, the broad-line region (BLR), and the dusty torus, is central to constraining the emission mechanisms \citep{Dermer1992, Sikora2009}.

In the radio regime, BZQs have classically been defined by flat spectra, with spectral indices $\alpha \leq 0.5$ between 1.4 and 4.85 GHz \citep{Gu2011}. The spectral index, defined as $S_{\nu} \propto \nu^{-\alpha}$, has long served as a diagnostic tool since the early works of \citet{Kellermann1969} and \citet{Condon1984}. \citet{Park2013}, using simultaneous high-frequency (22--43 GHz) data less affected by absorption, refined this scheme into three categories: steep ($\alpha \geq 0.5$), flat ($0 \leq \alpha < 0.5$), and inverted ($\alpha < 0$). More recently, other studies have estimated the spectral index at lower frequencies, specifically in the 1.4--3 GHz range \citep{Delvecchio2022}.

The observed radio spectra primarily reflect synchrotron self-absorption in compact jet regions, with additional curvature or convex shapes in cases where free-free absorption is significant \citep{Pacholczyk1973, Rybicki1979, Kameno2000, Kadler2004}.

With the growing availability of broadband and multi-epoch radio observations, simple index-based descriptions of radio spectra can be strongly affected by observational uncertainties and heterogeneous data coverage. This motivates careful reassessment of how radio spectral properties are characterized and interpreted in practice. The present work addresses this issue from an observational and phenomenological perspective by performing a systematic reassessment of BZQ radio spectra and introducing a statistical treatment that explicitly incorporates measurement uncertainties in spectral indices.


\section{SAMPLE}
\label{sec:Sample}

\subsection{Initial Selection}

We selected all sources classified as BZQs from the fifth edition of the Roma-BZCAT (5BZCAT; \citealt{Massaro2015}) and compiled their optical spectra from the sixteenth data release of the Sloan Digital Sky Survey (SDSS DR16; \citealt{Lyke2020}). After detailed visual inspection, we excluded spectra of poor quality, sources lacking broad emission lines, and objects with uncertain identifications. The resulting sample consists of 610 BZQs with at least one well-defined broad emission line, typically H$\beta$ $\lambda$4861\AA, Mg II $\lambda$2798\AA, or C IV $\lambda$1549\AA, depending on spectral coverage and redshift, spanning a redshift range of 0.11--5.28. The Roma-BZCAT is adopted as a starting point to construct a large and homogeneous sample of quasar-type blazars, while the designation “BZQ” is used throughout this work as a phenomenological label for sources spectroscopically confirmed as quasars. The complete multiwavelength catalog will be presented in a forthcoming paper (Guerrero-González et al., in preparation).

This initial selection ensured a large, homogeneous set of BZQs suitable for systematic verification of their blazar nature.

\subsection{Verification of the Blazar Nature}

Although the 5BZCAT catalog is specifically devoted to blazars, we assessed whether the selected BZQs display observational properties consistent with blazar-like behavior. This assessment combines optical spectroscopy, optical variability, and radio morphology, while recognizing that these diagnostics do not all carry the same weight in the classification scheme described below.

The presence of broad emission lines in the optical spectra was ensured by our initial selection and confirmed through visual inspection of SDSS data. We then used the detection of flare-like variability in optical light curves as the main observational indicator of Doppler-boosted synchrotron emission from relativistic jets aligned close to our line of sight \citep{Sher1968}. To test this, we analyzed $g$-band and $r$-band light curves from the Zwicky Transient Facility (ZTF; \citealt{Masci2019, Bellm2019}), constructed using forced photometry at the known position of each source \citep{Masci2023}. This approach provides consistent flux measurements across all epochs, including low signal-to-noise observations, and reduces artifacts associated with marginal detections in standard Data Release products. The resulting light curves span nearly eight years of systematic optical monitoring and are available for 98\% of the sample.

At this stage, flare-like variability is identified using a quantitative and reproducible procedure. As a first step, for each light curve we search for local maxima as preliminary flare candidates. This identification is performed using a dedicated Python routine based on the \texttt{signal.find\_peaks} function from the SciPy package, applying controlled thresholds in peak height, prominence, and minimum temporal separation to avoid spurious detections due to noise or irregular sampling.

For each light curve and for each photometric band independently, we define a single quiescent reference level $QB$ following \citet{Meyer2019}, which characterizes the baseline flux state of the source outside flare episodes. This reference level is used to identify significant flux enhancements above the quiescent state. Candidate peaks are required to exceed $QB$, so that small-amplitude fluctuations around the baseline are not automatically interpreted as flare-like events.

For each retained candidate peak, we fit a linear trend to $\ln F(t)$ within a backward time window $[t_{\rm peak}-W,t_{\rm peak}]$, adopting $W=100$ days. This window was chosen to test variability on timescales comparable to $\sim80$ days while retaining enough data points for statistically meaningful fits in irregularly sampled light curves. From the fitted logarithmic growth rate $m$, we compute the characteristic timescale associated with a flux increase by a factor $A$ as

\begin{equation}
\Delta t_A = \frac{\ln(A)}{m}.
\end{equation}

The literature reports optical variability episodes with different amplitudes and characteristic timescales. For instance, \citet{Gaur2012} describe faster brightening events with larger logarithmic growth rates, whereas \citet{Hayashida2015} report a systematic flux increase by a factor of $\sim4$ over a period of $\sim80$ days for the blazar 3C~279. From a methodological perspective, adopting the \citet{Hayashida2015} case corresponds to using a lower limiting growth rate, and therefore to a conservative and less restrictive definition of flare-like events. Accordingly, adopting $A=4$, we define

\begin{equation}
\Delta t_4 = \frac{\ln(4)}{m}.
\end{equation}

A candidate event is classified as flare-like only if it satisfies two conditions in at least one of the available ZTF optical bands ($g$ or $r$): (i) a statistically significant positive logarithmic slope ($p$-value $<0.05$), and (ii) a characteristic timescale $\Delta t_4 \leq 80$ days. With the adopted amplitude and timescale, this is equivalent to requiring a minimum logarithmic growth rate of $m_{\rm thr}=\ln(4)/80 \simeq 0.017\,{\rm day}^{-1}$. Events with positive but much smaller slopes correspond to slower, long-term variability rather than flare-like behavior under our adopted definition. Similarly, negative slopes, statistically insignificant fits, or poorly sampled windows do not yield a valid flare classification.

\begin{figure}[ht!]
\centering
\includegraphics[width=\columnwidth]{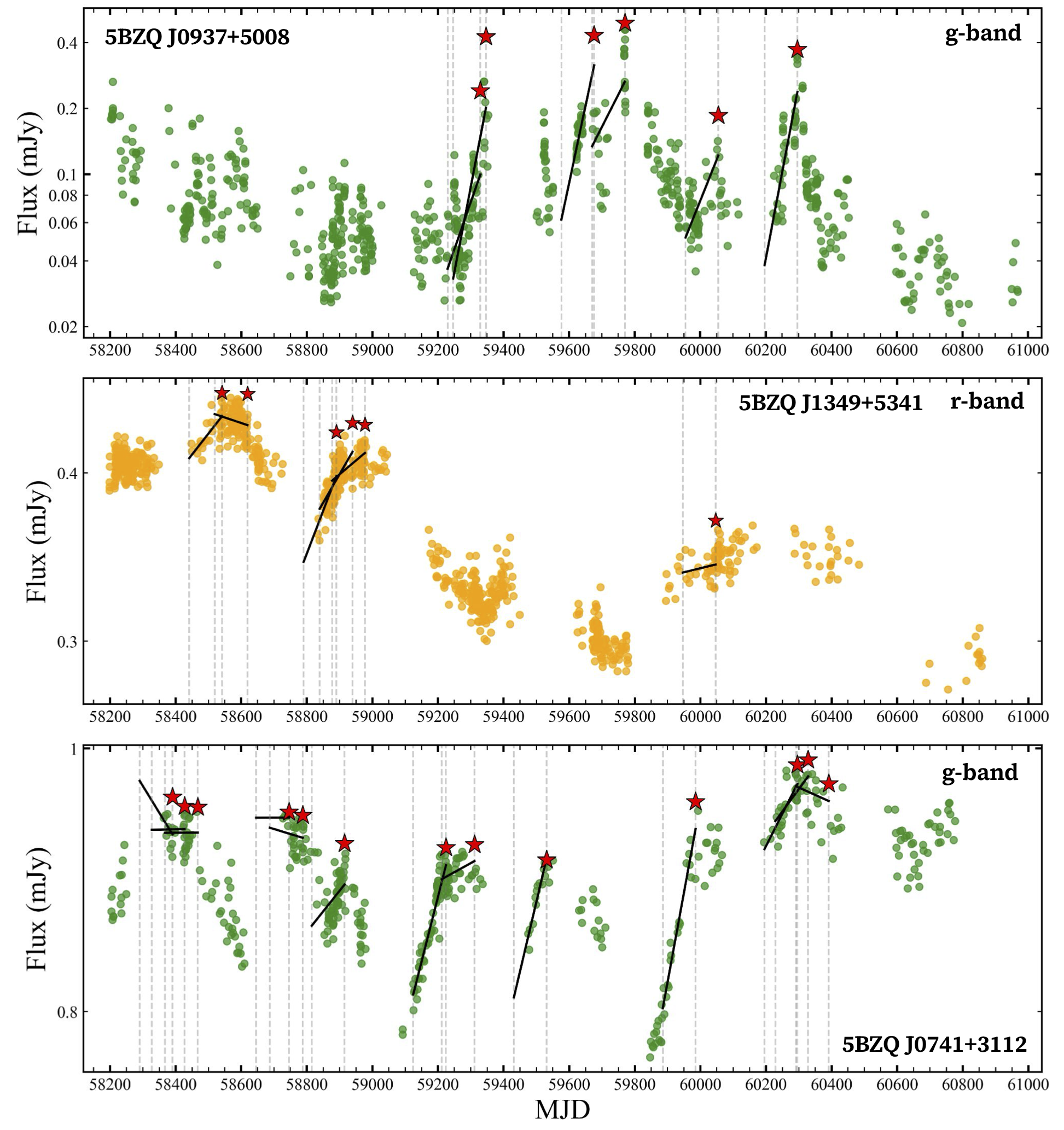}
\caption{Representative examples of the quantitative optical flare-identification procedure applied to ZTF forced-photometry light curves. The top, middle, and bottom panels show 5BZQ~J0937+5008 ($g$ band), 5BZQ~J1349+5341 ($r$ band), and 5BZQ~J0741+3112 ($g$ band), respectively, corresponding to the representative sources discussed throughout the manuscript. All flux axes are plotted on a logarithmic scale. Red stars mark candidate peaks identified by the automated procedure, while black lines show the linear fits performed on $\ln F(t)$ within the backward time window of $W=100$ days ending at each peak. Vertical dashed lines delimit the fitting windows. From these fits, we derive the logarithmic growth rate $m$, the characteristic timescale $\Delta t_4$, and the associated statistical significance. Only candidate peaks satisfying both fit $p$-value $<0.05$ and $\Delta t_4 \leq 80$ days are classified as flare-like under the adopted criterion.}
\label{fig:optical_flare_examples}
\end{figure}

The application of this procedure is illustrated in Fig.~\ref{fig:optical_flare_examples} and the results are summarized in Table~\ref{tab:optical_flare_examples}. For 5BZQ~J0937+5008, two candidate peaks satisfy the adopted flare-like criterion, while four additional peaks do not; one of these non-flaring peaks is included in the table for comparison. The detection of at least one statistically significant and sufficiently rapid flux increase is taken as evidence of flare-like optical variability, consistent with Doppler-boosted emission from a compact jet region oriented close to the line of sight. In contrast, for 5BZQ~J1349+5341 and 5BZQ~J0741+3112, none of the candidate peaks satisfies the adopted criterion, although representative non-flaring peaks are shown in the table. These examples show that a source may display visible optical variability and still not satisfy the quantitative flare criterion. In such cases, the source is considered variable in a broad sense, but not flare-like for the purposes of the present phenomenological classification. A dedicated and broader variability characterization for the full sample will be presented in Paper~II of this series; here, flare detection is used only as a conservative classification filter.

\begin{table*}[ht!]
\centering
\caption{Examples of the peak-analysis results for the sources shown in Fig.~\ref{fig:optical_flare_examples}. A candidate peak is classified as flare-like only when it simultaneously satisfies fit $p$-value $<0.05$ and $\Delta t_4 \leq 80$ days in at least one band.}
\label{tab:optical_flare_examples}
\begin{tabular}{c c c c c c c}
\hline
Source & Band & MJD$_{\rm peak}$ & $m$ (day$^{-1}$) & $\Delta t_4$ (days) & $p$-value & Interpretation \\
\hline
5BZQ J0937+5008 & $g$ & 59346.33 & $1.80\times10^{-2}$ & 76.81 & $<10^{-16}$ & Flare \\
 &  & 59770.17 & $6.81\times10^{-3}$ & 203.59 & $0.0021$ & Non-flare \\
 &  & 60296.34 & $1.83\times10^{-2}$ & 75.82 & $2.21\times10^{-10}$ & Flare \\
5BZQ J1349+5341 & $r$ & 58890.47 & $1.47\times10^{-3}$ & 944.82 & $6.73\times10^{-9}$ & Non-flare \\
 &  & 58939.40 & $9.72\times10^{-4}$ & 1425.53 & $2.22\times10^{-15}$ & Non-flare \\
 &  & 58977.25 & $4.73\times10^{-4}$ & 2933.71 & $1.95\times10^{-12}$ & Non-flare \\
5BZQ J0741+3112 & $g$ & 59224.43 & $1.10\times10^{-3}$ & 1263.39 & $<10^{-16}$ & Non-flare \\
 &  & 59531.38 & $1.18\times10^{-3}$ & 1172.23 & $1.75\times10^{-8}$ & Non-flare \\
 &  & 59985.35 & $1.52\times10^{-3}$ & 910.86 & $1.60\times10^{-10}$ & Non-flare \\
\hline
\end{tabular}
\end{table*}

Under this classification scheme, sources without clear flare-like variability cannot be classified as confirmed blazars. Then, radio morphology was used as complementary evidence to then distinguish between possible and non-confirmed blazars. We inspected high-resolution data at multiple frequencies: 1.4 GHz images from FIRST\footnote{\url{https://sundog.stsci.edu/cgi-bin/searchfirst}} (5$^{\prime\prime}$ resolution; \citet{Becker1995}, available for 91\% of the sample), 144 MHz images from LOFAR LoTSS\footnote{\url{https://lofar-surveys.org/dr2_release.html}} (6$^{\prime\prime}$ resolution; \citet{vanHaarlem2013}; \citet{Shimwell2022}, 57\%), and VLBI data at 0.1--1 mas resolution from the Astrogeo archive\footnote{\url{http://astrogeo.org/vlbi_images/}} (\citet{Petrov2009}; \citet{Pushkarev2012}; \citet{Piner2012}, 96\%). As expected for blazars, the typical morphology corresponds to compact cores or one-sided jets, with counter-jets suppressed by relativistic beaming. In contrast, symmetric double-lobed structures are consistent with larger viewing angles and jets not aligned with the line of sight. We therefore flagged as Non-Confirmed BZQ candidates, those sources showing bilateral morphology in LOFAR images, or in FIRST when LOFAR coverage was unavailable.

Some sources displayed discrepancies across spatial scales. For instance, a subset exhibited bilateral morphology at 144 MHz while simultaneously showing flare-like optical variability. Others revealed significant misalignments between parsec-scale (VLBI) and kiloparsec-scale (FIRST/LOFAR) jet orientations, a phenomenon also noted by \citet{Kharb2010}. Such behavior suggests that while kpc-scale jets may resemble those of radio galaxies, the inner cores can retain a blazar-like nature, i.e. the pc-scale jet is aligned towards our line of sight. This can potentially occur due to intrinsic nuclear processes or environmental influences (e.g., binary black holes or mergers) that alter jet orientation. To explore this, we measured jet position angles on both scales, defining angles relative to the west--east axis (0$^{\circ}$). For kiloparsec scales, the angle was traced from the core to the brightest lobe in FIRST/LOFAR images; for parsec scales, from the VLBI core to the most prominent non-core component, representing the jet axis. Sources appearing fully compact or with diffuse halos lacking a clear jet were not assigned an angle. Our measurements were cross-checked against the automated determinations of \citet{Plavin2022} for 9220 AGNs across 1.4--86 GHz, showing good consistency for overlapping objects.

This verification step allowed us to identify sources whose observed properties are consistent with blazar-like behavior and to distinguish them from cases where the available data provide insufficient or ambiguous evidence, forming the basis for a phenomenological consistency assessment of the sample.

\subsection{Phenomenological consistency assessment of BZQs}

Based on these diagnostics, we grouped the sources into three categories according to their observational consistency with blazar-like behavior. We emphasize that the categories are not defined by a simple count of equally weighted criteria: broad emission lines are part of the initial sample definition, optical flare-like variability provides the strongest and sufficient evidence for a Doppler-boosted jet, and radio morphology is used only as complementary information.

\begin{itemize}
    \item Confirmed BZQ: Sources that exhibit broad emission lines and clear flare-like optical variability, regardless of their radio morphology.

    \item Possible BZQ: Sources showing gradual or low-amplitude optical variability, but lacking clear flare-like activity under the adopted criterion. These sources display compact cores or one-sided radio jets.

    \item Non-Confirmed BZQ: Sources lacking clear flare-like optical variability and showing radio morphologies that do not provide supporting evidence for blazar-like behavior, characterized by symmetric or double-lobed structures. This category does not imply that the sources are non-blazars, but rather that the available data do not provide sufficient evidence to confirm their blazar-like behavior under the criteria adopted in this work.
\end{itemize}

We note that the absence of detected flare-like variability within the available ZTF temporal baseline is not, by itself, evidence against a blazar nature. Blazars can undergo extended periods of low activity or quiescence, during which no prominent flares are observed over timescales of several years (e.g., \citealt{Chavushyan2020}; \citealt{Fernandes2020}; \citealt{Roy2021}; \citealt{Thekkoth2024}). Therefore, the Non-Confirmed BZQ label should be interpreted strictly as an observational and phenomenological designation, derived from the available data, not as a definitive physical classification.

\begin{figure}[ht!]
\centering
\includegraphics[width=\columnwidth]{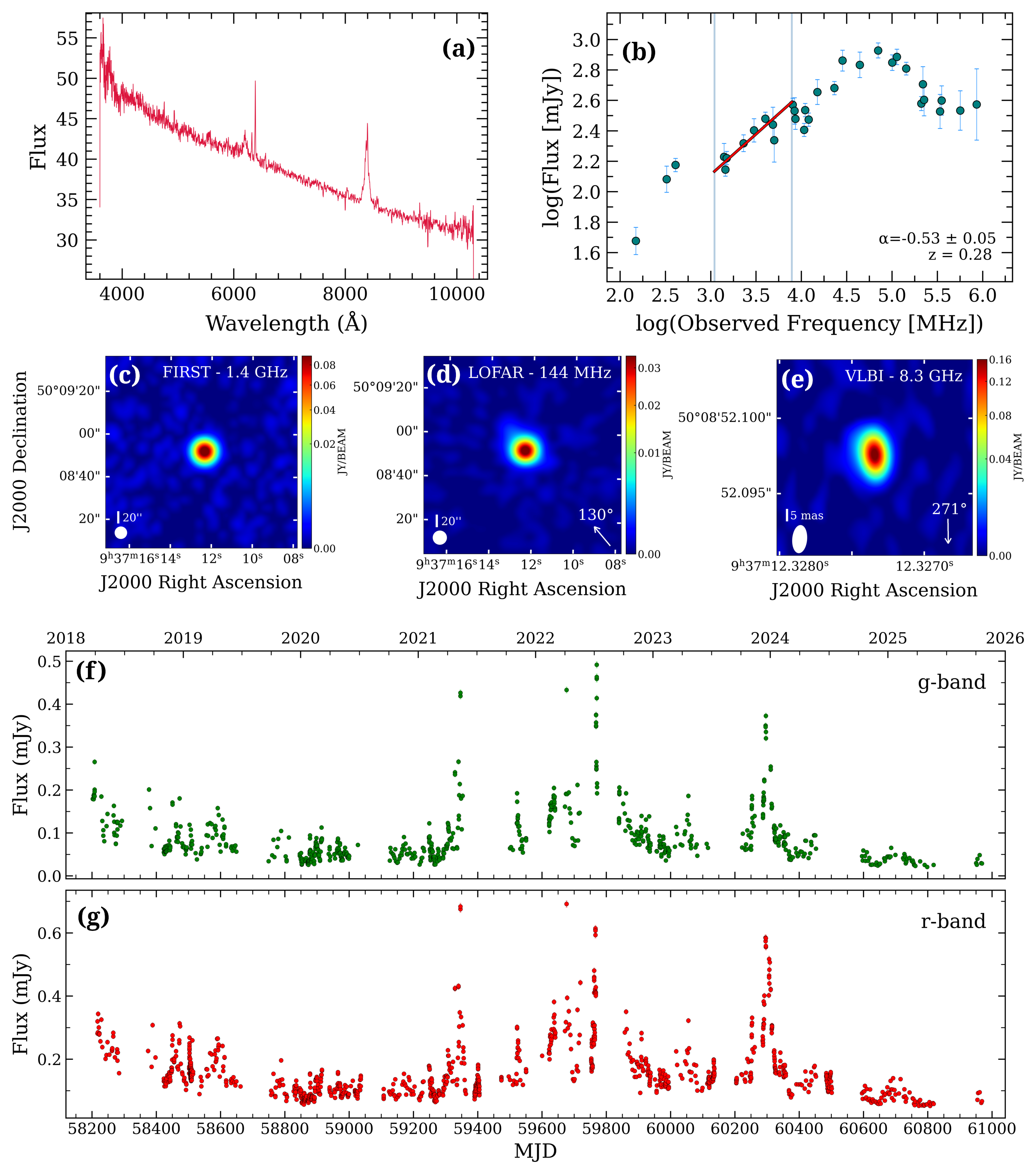}
\caption{Multi-wavelength characterization of 5BZQ J0937+5008 (Confirmed BZQ). (a) SDSS optical spectrum displaying prominent broad emission lines characteristic of quasar-type sources. (b) Broadband radio spectrum constructed from multi-frequency flux-density measurements obtained from CATS and the SED Builder. The solid red line indicates the best-fit spectral slope within the selected rest-frame frequency interval (1.4–10 GHz), while the vertical blue lines mark the boundaries of this interval. The derived spectral index $\alpha$, its uncertainty, and the source redshift are indicated in the panel. (c–e) Radio images at different frequencies and spatial scales: FIRST at 1.4 GHz, LOFAR/LoTSS at 144 MHz, and VLBI at 8.7 GHz, respectively. Color bars indicate flux density in mJy beam$^{-1}$, and synthesized beams are shown in the lower corners of each panel. (f–g) ZTF optical light curves in the g and r bands, spanning nearly eight years of monitoring. The source exhibits clear flare-like variability in both bands, consistent with Doppler-boosted jet emission. The source satisfies the spectroscopic requirement and shows clear flare-like optical variability, which is the main and sufficient evidence for confirmed blazar-like behavior in our framework. Its compact/one-sided radio morphology provides additional supporting evidence for its classification as a Confirmed BZQ.}
\label{fig:ConfirmedBZQ}
\end{figure}

\begin{figure}[ht!]
\centering
\includegraphics[width=\columnwidth]{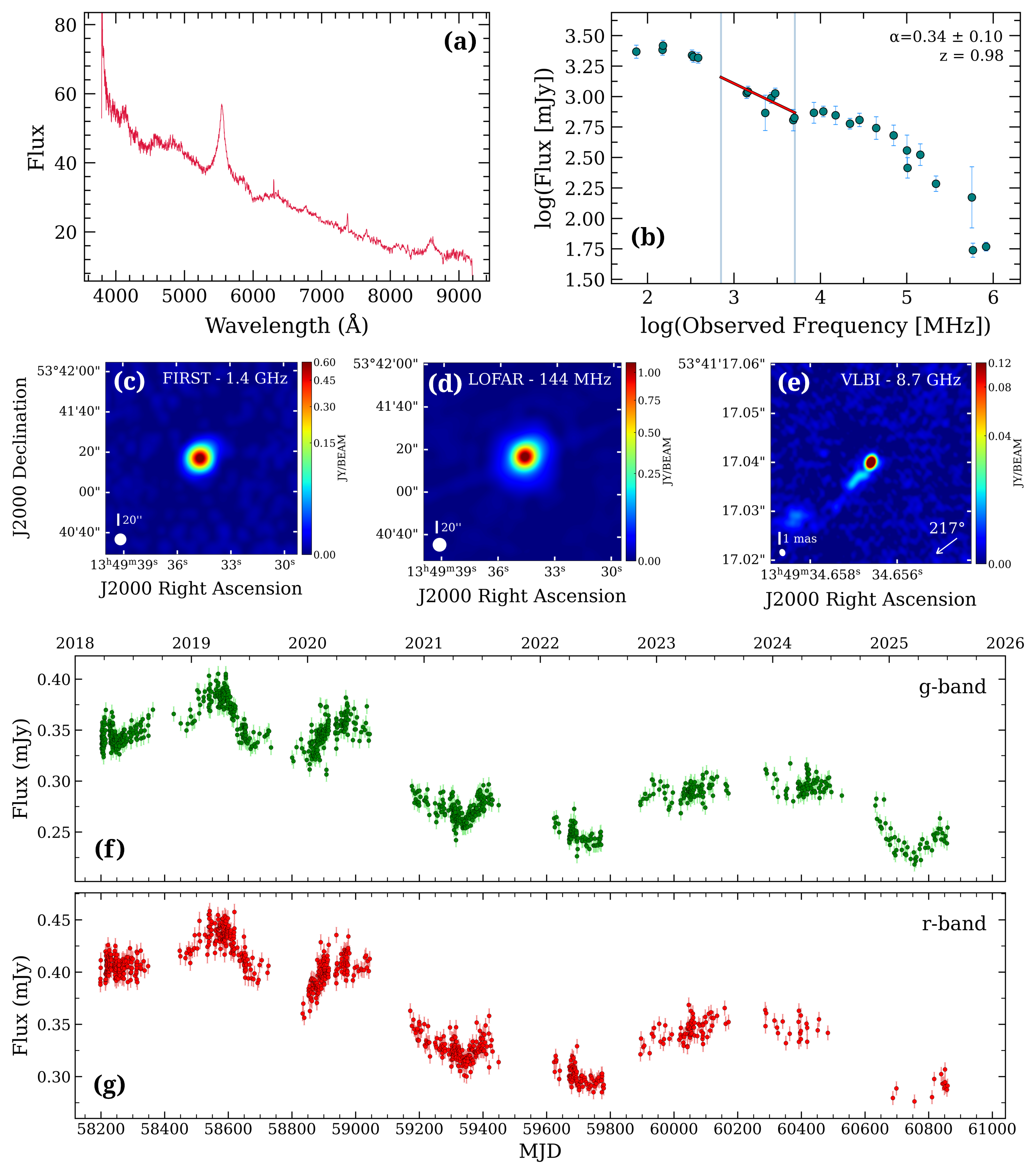}
\caption{Multi-wavelength characterization of 5BZQ J1349+5341 (Possible BZQ). Same panels as in Figure~\ref{fig:ConfirmedBZQ}. This source exhibits broad emission lines and compact radio morphology but shows no clear flare-like optical variability during the observed period, consistent with its classification as a Possible BZQ.}
\label{fig:PossibleBZQ}
\end{figure}

\begin{figure}[ht!]
\centering
\includegraphics[width=\columnwidth]{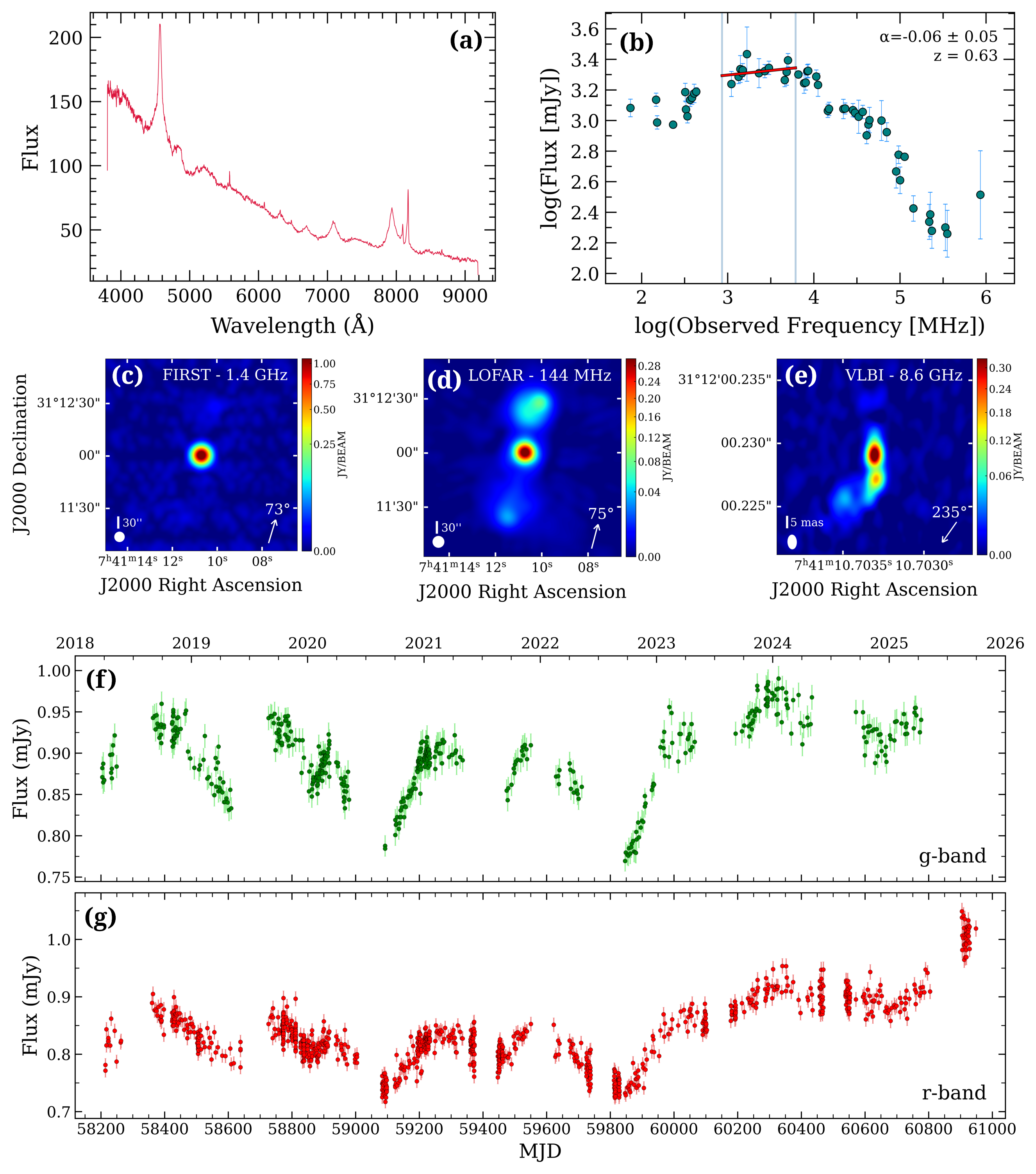}
\caption{Multi-wavelength characterization of 5BZQ J0741+3112 (Non-Confirmed BZQ). Same panels as in Figure~\ref{fig:ConfirmedBZQ} This source shows a symmetric double-lobed radio morphology. Although its optical light curves may display variability in a broad sense, the candidate peaks do not satisfy the adopted quantitative flare-like criterion. Under the criteria adopted in this work, the available data therefore provide insufficient evidence for a Confirmed BZQ classification.}
\label{fig:NonBZQ}
\end{figure}

Applying this scheme, 33\% of the sample were classified as Confirmed BZQs, 54\% as Possible BZQs, and 13\% as Non-Confirmed BZQs. The latter category identifies sources whose available optical and radio data do not provide sufficient evidence for blazar-like behavior under the adopted criteria, highlighting the role of observational limitations and the need for additional constraints to achieve a more robust and internally consistent classification. A refined catalog, including these classifications and further physical properties, will be presented in a forthcoming publication. Representative examples of each category are shown in Figures~\ref{fig:ConfirmedBZQ}--\ref{fig:NonBZQ}.

Having established the classification framework, we next turned to the broadband radio spectra, which are the foundation for our subsequent spectral analysis.

\subsection{Radio Spectra Data}

For the 610 sources in our sample, we constructed broadband radio spectra by combining flux-density measurements from the CATS catalog (\citealt{Verkhodanov1997}) with additional data retrieved via the SED Builder tool (\citealt{Stratta2011}). This approach provided continuous frequency coverage from a few MHz to several hundred GHz, forming the basis for the spectral analyses presented in this work. The full set of original radio spectra for the 610 sources is available in the Zenodo dataset\footnote{\url{https://zenodo.org/records/20145648}} \citep{GuerreroGonzalez2026}. 


\section{RADIO SPECTRA CLASSIFICATION}
\label{sec:Radio Spectra Classification}

BZQs have traditionally been identified by their flat radio spectra, but given the phenomenological consistency assessment presented in Section~\ref{sec:Sample}, it is necessary to verify whether our refined sample consistently satisfies this property. To this end, we adopted the classification framework of \citet{Park2013}, which categorizes sources according to their spectral shapes within a defined rest-frame frequency interval.

\begin{figure}[ht!]
\centering
\includegraphics[width=\columnwidth]{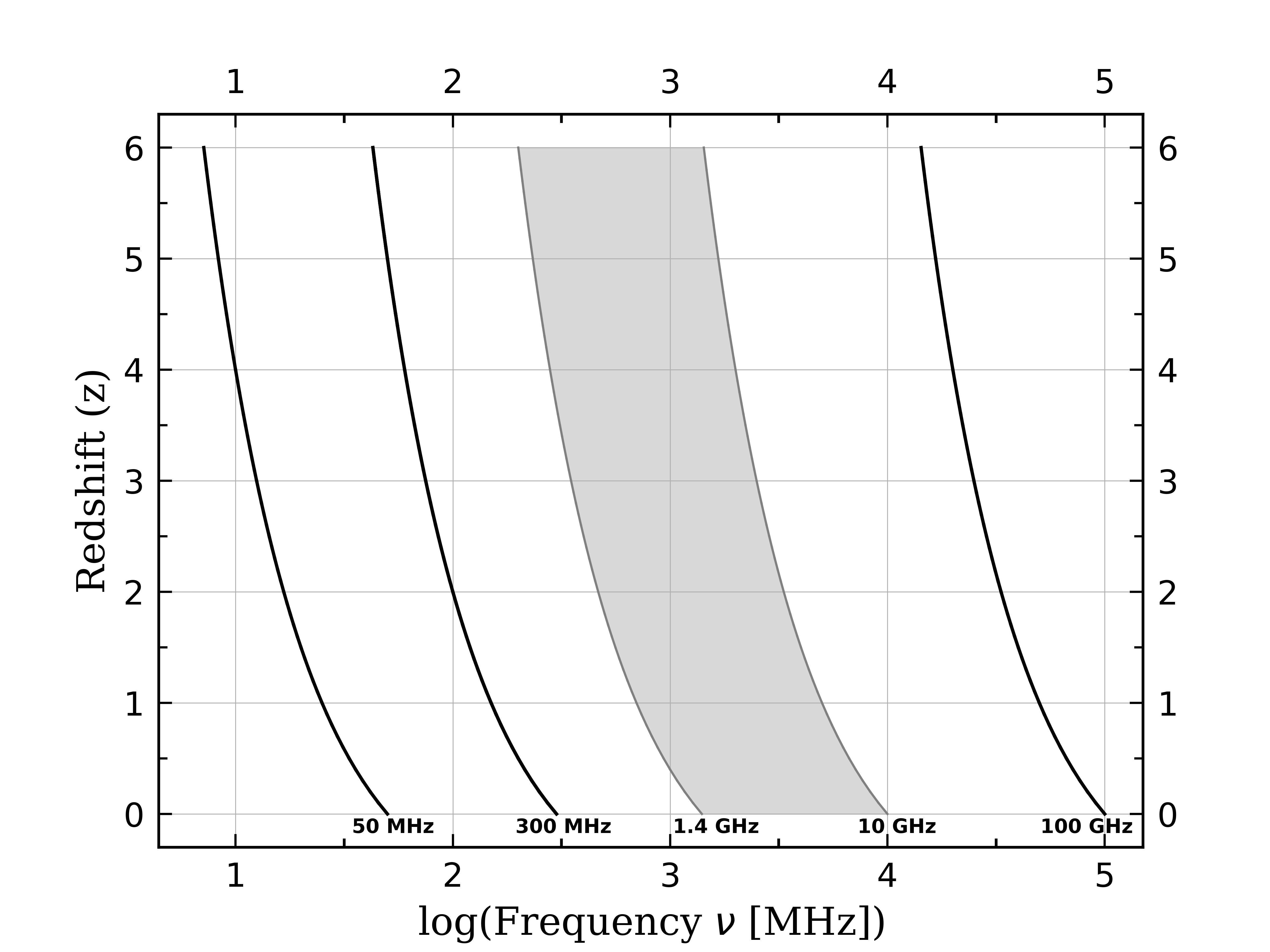} 
\caption{Observed frequency as a function of redshift for fixed rest-frame frequencies. The black curves show the projection of several rest-frame frequencies (50 MHz, 300 MHz, 1.4 GHz, 10 GHz, and 100 GHz) into the observer’s frame. The shaded region marks the interval adopted for spectral classification (1.4–10 GHz in the rest frame), within which the linear fits were performed.}
\label{fig:zFreq}
\end{figure}

For our analysis, we selected the rest-frame frequency range 1.4--10 GHz. This interval was chosen to focus on the spectral region where radio emission is least affected by low-frequency curvature associated with extended kpc-scale structures and by high-frequency steepening or variability-driven effects. Within this range, the majority of sources are well described by a single power-law, facilitating a more homogeneous comparison across redshifts. This choice minimizes the influence of additional components at very low or high frequencies and allows for consistent cross-redshift comparisons. For each source, we computed the observed frequency range corresponding to 1.4--10 GHz in the rest frame, according to its redshift. This approach shifts the interval systematically towards lower frequencies without recalibrating the full spectrum, thereby preserving the overall spectral shape. Figure~\ref{fig:zFreq} provides a schematic illustration of how the selected rest-frame interval maps onto observed frequencies as a function of redshift, and motivates its adoption by highlighting the increased complexity and curvature commonly encountered outside this range.

Within this interval, we fit each spectrum in logarithmic space to derive the radio spectral index $\alpha$, defined as $S_{\nu} \propto \nu^{-\alpha}$. Following the conventional scheme \citep{Park2013}, spectra with $0 \leq \alpha < 0.5$ are considered flat. This parameter forms the basis of our radio spectral classification.

\subsection{Methodology}

Flux-density data were compiled from the CATS and SED Builder catalogs. The values, originally provided in linear scale, were converted to logarithmic units for both frequency and flux, with errors propagated accordingly. Data points with uncertainties $>$100\% were excluded. For points with unrealistically small errors ($<$5\%), a conservative 10\% error was adopted to avoid bias from under-reported uncertainties.

Duplicate entries at the same (log $\nu$, log $S_{\nu}$) were resolved by retaining the measurement with the largest error, ensuring conservative estimates. To mitigate variability-induced outliers, we applied a block-based filter over 0.3 dex in log\,$\nu$, starting the first block on the lowest frequency on the data: within each block we computed the mean and standard deviation of log\,$S_{\nu}$ and retained only measurements with log\,$S_{\nu} \in [\mu \pm 1\sigma]$ (a deliberately strict threshold to enforce homogeneity). This procedure removed variability outliers from the spectra, while single-point blocks were kept unchanged. By construction, this frequency-binning and filtering strategy is intended to suppress transient high- and low-activity states and to approximate the typical, time-averaged radio spectral behavior of each source, rather than instantaneous spectral states. We note, however, that the use of non-simultaneous archival data may still introduce residual variability effects in individual cases, although these are not expected to significantly affect the statistical properties of the sample as a whole.

Because the spectra span several orders of magnitude in frequency, the binning was performed in logarithmic frequency space. This choice preserves a constant relative frequency resolution across the spectrum ($\Delta\nu/\nu$ is constant). Although larger bins could reduce the impact of variability-driven scatter, they would also increase the risk of mixing physically distinct spectral regions and smoothing out genuine spectral curvature.

We then consolidated measurements by grouping values with nearly identical frequencies ($\Delta\log\nu \leq 0.01$). Each group was replaced by a single point whose frequency corresponds to the group mean, whose flux is the error-weighted mean (weights $1/\sigma^{2}$), and whose uncertainty was set to either the group’s (unweighted) standard deviation or the highest individual error, depending on which value is higher (to be conservative). This procedure reduced redundancy and yielded a cleaner dataset for spectral fitting, since having multiple data points very close in frequency would skew the fitting.

The spectral indices were then derived by performing linear fits in the 1.4--10 GHz rest-frame interval using the MagicPlot software.\footnote{\url{https://magicplot.com/}} Each fit provided slope ($\alpha$), intercept, and associated errors using an error-weighted least-squares algorithm. For spectra with sparse coverage in the classification interval, the fit was allowed to extend marginally into adjacent frequencies if the trend remained coherent; otherwise, it was restricted to the defined range. Manual inspection was carried out for every spectrum, excluding isolated outliers with large errors (often attributable to intrinsic variability within small frequency ranges). For spectra exhibiting breaks within the classification range, the fit was performed on the subset with greater data density and clearer trend. When the fitting interval was defined by only two points, but marginal data were available at nearby frequencies, the fit was extended following the consistent trend of those marginal points. In extreme cases, these marginal data were explicitly incorporated to obtain a more stable and physically consistent fit.

\begin{figure}[ht!]
\centering
\includegraphics[width=\columnwidth]{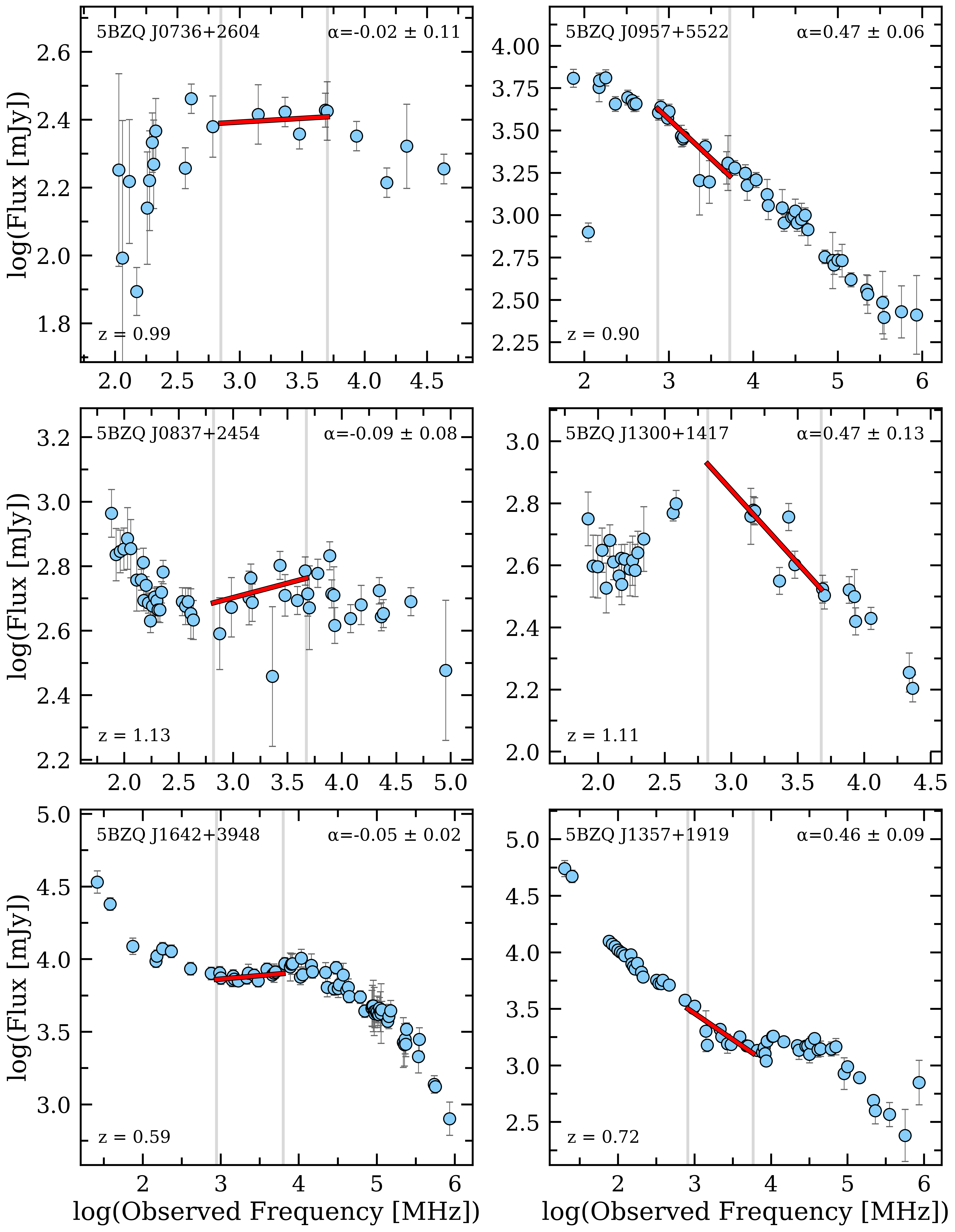}
\caption{Examples of radio spectra after the cleaning procedure. Blue points show flux-density measurements with associated uncertainties, while red lines represent the error-weighted linear fits performed within the 1.4–10 GHz rest-frame interval. The vertical gray lines mark the boundaries of this interval, whose position in the observed frame depends on source redshift (see Figure~\ref{fig:zFreq}). Each panel indicates the source name, redshift, and derived spectral index ($\alpha$) with its uncertainty. The complete set of spectra for all sources is available in the Zenodo dataset \citep{GuerreroGonzalez2026}.}
\label{fig:RadioSpecExamples}
\end{figure}

Examples of final radio spectra are shown in Figure~\ref{fig:RadioSpecExamples}. The complete atlas of cleaned and fitted radio spectra is provided in the Zenodo dataset \citep{GuerreroGonzalez2026}. This methodology ensured that the derived spectral indices are robust against measurement errors, variability, and data inhomogeneities, providing a solid foundation for re-examining the classical Flat-spectrum criterion.

\subsection{Proposed Classification}

Application of the \citet{Park2013} scheme to our cleaned spectra revealed important inconsistencies. Figure~\ref{fig:RadioSpecExamples} illustrates two representative cases:

\begin{itemize}
    \item Left-hand panels: spectra with negative but near-zero indices (e.g., 5BZQ J0736+2604, $\alpha = -0.02 \pm 0.11$). According to the standard criterion, these would be classified as inverted ($\alpha < 0$), yet their spectral shapes clearly correspond to nearly flat spectra within uncertainties.
    
    \item Right-hand panels: spectra with $\alpha$ values formally in the flat range ($0 < \alpha < 0.5$), but with slopes visually consistent with steep spectra. Accounting for errors, these cases could be equally well classified as steep.
\end{itemize}

\begin{figure}[ht!]
\centering
\includegraphics[width=\columnwidth]{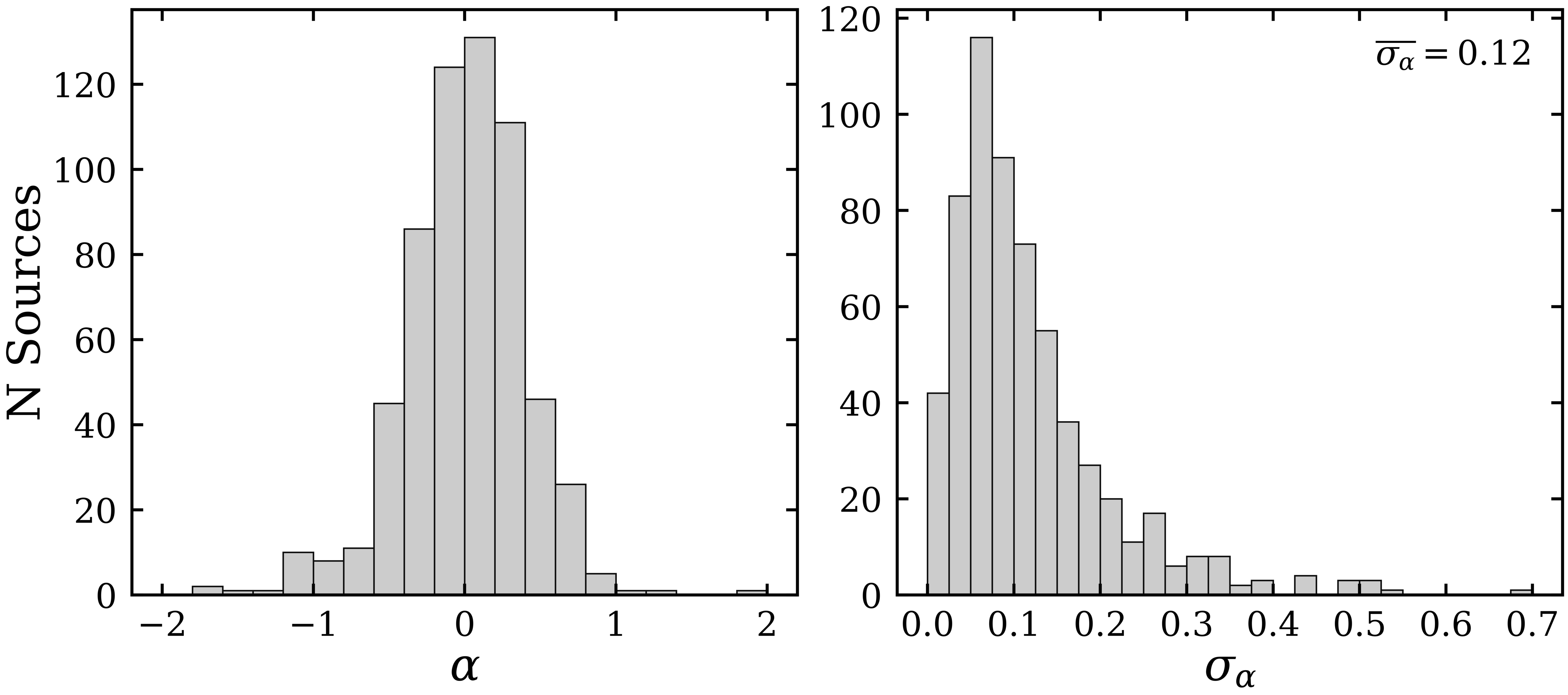} 
\caption{Distribution of radio spectral indices and their associated uncertainties for the 610 sources in the sample. Left panel: histogram of spectral indices ($\alpha_i$). Right panel: histogram of individual uncertainties ($\sigma_{\alpha,i}$). In the classification scheme, these uncertainties are used on a source-by-source basis to evaluate whether each spectral index is consistent with $\alpha=0$.}
\label{fig:Alphas}
\end{figure}

Such ambiguities arise because the classical thresholds do not consider the statistical significance of the spectral-index measurement relative to its uncertainty. Figure~\ref{fig:Alphas} shows the distributions of the measured spectral indices $\alpha_i$ and their individual uncertainties $\sigma_{\alpha,i}$. Since the uncertainty distribution is not strictly Gaussian and varies from source to source, a single uncertainty value for the entire sample would not adequately represent the measurement precision of individual objects. We therefore adopt a source-by-source classification scheme based on the individual uncertainty $\sigma_{\alpha,i}$ associated with each spectral index.

For each source, we quantify the significance of the deviation from a flat spectrum as $|\alpha_i|/\sigma_{\alpha,i}$. We then classify each source according to whether its spectral index is consistent with $\alpha=0$ within an adopted significance threshold. In this work, we adopt a $2\sigma$ criterion:

\begin{itemize}
    \item Flat: $|\alpha_i| \leq 2\sigma_{\alpha,i}$
    \item Steep: $\alpha_i > 2\sigma_{\alpha,i}$
    \item Inverted: $\alpha_i < -2\sigma_{\alpha,i}$
\end{itemize}

This source-by-source criterion explicitly incorporates the uncertainty associated with each spectral-index measurement. The choice of a $2\sigma$ threshold is motivated by the population-level goal of this study. A more conservative $3\sigma$ criterion would minimize false positives, but would also retain in the flat category sources that already show evidence for non-zero spectral indices at the $\sim95\%$ confidence level. For example, 5BZQ~J0126+2559 ($\alpha=0.32\pm0.13$, $2.5\sigma$) and 5BZQ~J0218--0923 ($\alpha=0.38\pm0.13$, $2.9\sigma$), together with 15.57\% of the sample, fall between the $2\sigma$ and $3\sigma$ limits. Classifying all such sources as flat would mask statistically meaningful deviations from $\alpha=0$ and artificially reduce the inferred fraction of non-flat spectra.

We therefore adopt the $2\sigma$ source-by-source criterion as a balance between false positives and false negatives, consistent with the phenomenological aim of this work. For transparency, we also show the classification obtained with an alternative $3\sigma$ threshold in Fig.~\ref{fig:three_sigma_classification}. As expected, the $3\sigma$ criterion increases the fraction of flat spectra in all three populations (only 15.6\% of the sample changed classification), but the qualitative conclusion remains unchanged: the BZQ population exhibits a broad range of radio spectral behavior, and not all sources are strictly flat within the observational uncertainties.

This scheme reclassifies the borderline cases in a more statistically consistent way: the left-hand spectra in Figure~\ref{fig:RadioSpecExamples} fall into the flat category, while the right-hand spectra are consistently identified as steep. Table~\ref{tab:Tabla1} presents examples of the spectral indices ($\alpha$) and their uncertainties, together with the classification according to the \citet{Park2013} scheme and the revised classification proposed in this work. The full catalog of results for the 610 sources is available in the Zenodo dataset \citep{GuerreroGonzalez2026}.

\begin{table}[ht]
\caption{Examples of spectral indices ($\alpha$) with uncertainties and classifications from 1.4--10\, GHz fits.}

\centering
\small
\begin{tabular}{c c c c c}
\hline
Source 5BZQ & $\alpha^{a}$ & $\sigma_{\alpha}^{b}$ & Park et al. (2013)$^{c}$ & This Work$^{d}$ \\
\hline
J0006+2422 & $-$0.05 & $\pm$0.08 & Inverted & Flat \\
J0029+0554 & $-$0.24 & $\pm$0.11 & Inverted & Inverted \\
J0126+2559 & +0.32 & $\pm$0.13 & Flat & Steep \\
J0151+2744 & +0.20 & $\pm$0.05 & Flat     & Steep \\
J0218$-$0923 & +0.38 & $\pm$0.13 & Flat & Steep \\
J0239+0416 & $-$0.02 & $\pm$0.09 & Inverted & Flat \\
J0808+4752 & +0.99 & $\pm$0.51 & Steep    & Flat \\
J0914+0245 & $-$0.34 & $\pm$0.08 & Inverted & Inverted \\
J0916+3854 & +0.43 & $\pm$0.07 & Flat     & Steep \\
J0920+4441 & $-$0.52 & $\pm$0.16 & Inverted & Inverted \\
J0921+6215 & +0.15 & $\pm$0.10 & Flat     & Flat \\
J0923+2815 & $-$0.26 & $\pm$0.21 & Inverted & Flat \\
\bottomrule
\multicolumn{5}{l}{\textbf{Notes.}}\\
\multicolumn{5}{l}{$^{a}$ Spectral index from the 1.4--10\,GHz rest-frame fit.}\\
\multicolumn{5}{l}{$^{b}$ $1\sigma$ uncertainty on $\alpha$.}\\
\multicolumn{5}{l}{$^{c}$ Classification according to Park et al. (2013)}\\
\multicolumn{5}{l}{$^{d}$ Classification adopted in this work.}\\
\end{tabular}
\label{tab:Tabla1}
\end{table}

Our results also challenge long-standing assumptions. For instance, early studies suggested a representative $\alpha \approx 1$ for AGN populations \citep{Kellermann1969, Punsly2005}. More recent long-term radio monitoring programs, such as F-GAMMA \citep{Angelakis2012} and MOJAVE \citep{Lister2005, Hovatta2014}, have since provided a more comprehensive view, showing that blazar radio spectra span a wide range of spectral indices and exhibit significant variability, rather than clustering around a single characteristic value. In this context, the distribution obtained here (Figure~\ref{fig:Alphas}) demonstrates that such a value does not characterize the average BZQ, underscoring the broad spectral diversity within this class.

This revised classification thus provides a more reliable framework for interpreting BZQ radio spectra, paving the way for the statistical analyses presented in the next section.


\section{FULL RADIO SPECTRUM CLASSIFICATION}
\label{sec:Full Radio Spectrum Classification}

Radio spectra in blazars are intrinsically time-dependent due to variability in the relativistic jet. In large-sample studies, it is common to construct broadband spectra from non-simultaneous archival data (e.g. \citealt{Boettcher2013}), since fully simultaneous observations over a broad frequency range are rarely available. In this context, the derived spectral indices and full-spectrum morphologies should be interpreted as effective, time-averaged descriptors of the typical spectral behavior of each source, rather than as specific instantaneous emission states. Variability can introduce artificial slopes or curvature when measurements obtained at different epochs correspond to different flux states. While our binning procedure mitigates the influence of extreme outliers and reduces the impact of scattered measurements, it cannot fully remove this effect.

To address this limitation, we visually inspected the full set of radio spectra to identify cases potentially affected by distinct non-simultaneous radio states. In this inspection, we considered as potentially affected cases only those spectra showing coherent, separated flux-density tracks over comparable frequency ranges, rather than isolated discrepant points or scatter compatible with the reported uncertainties. Under this conservative criterion, we identified 9 candidate cases, corresponding to only 1.5\% of the full sample. Therefore, the classifications presented below should be understood as phenomenological descriptions of the observed broadband spectra; individual cases affected by non-simultaneity should be interpreted with caution, although the statistical properties of the sample are expected to remain robust.

\begin{figure}[ht!]
\centering
\includegraphics[width=\columnwidth]{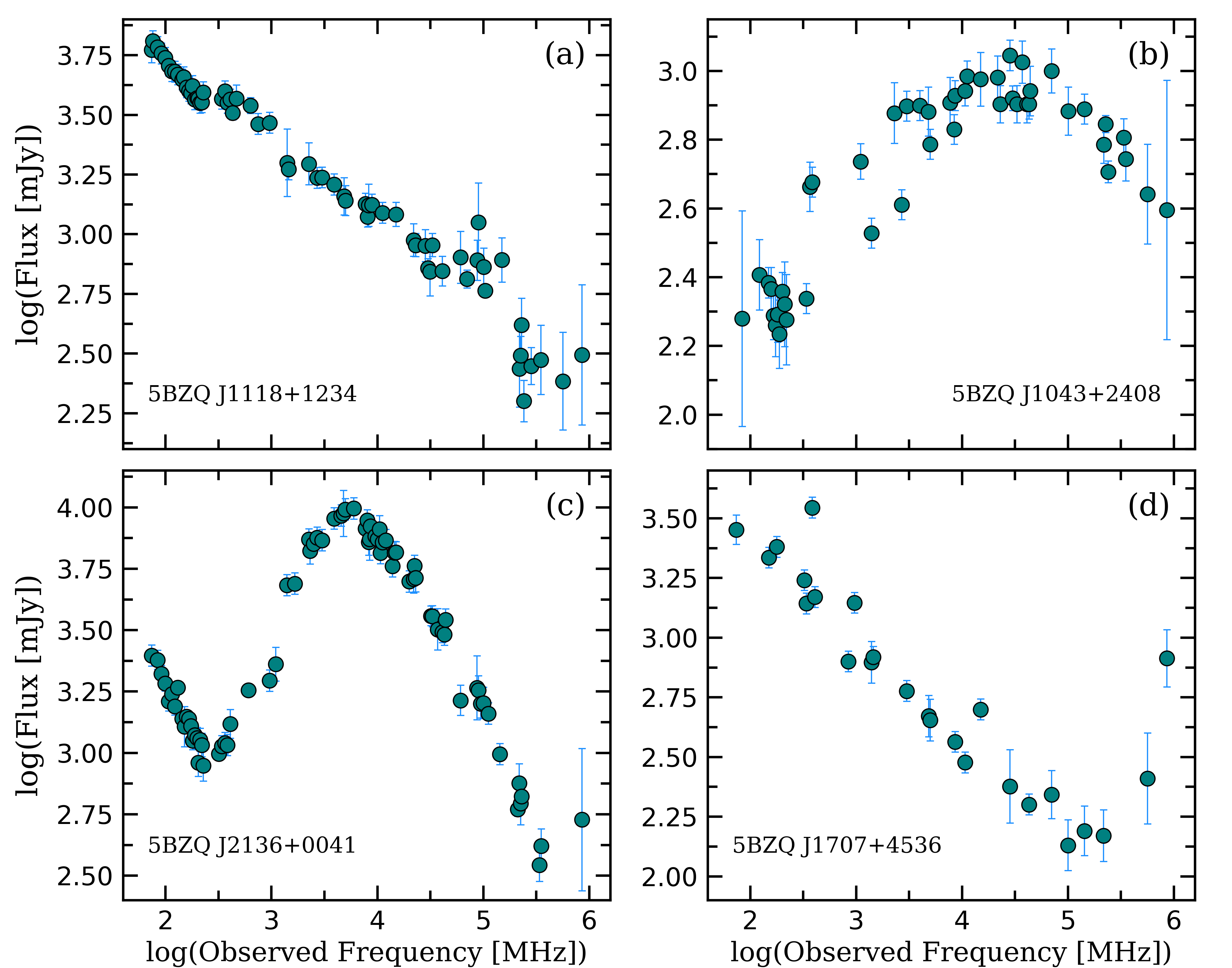} 
\caption{Representative examples of the full radio spectral morphologies identified in this work. (a) Power-law spectrum, consistent with optically thin synchrotron emission. (b) Peaked spectrum, shaped by synchrotron self-absorption at low frequencies. (c) Retriggered-peaked spectrum, produced by the superposition of multiple particle-injection episodes. (d) Inverted-peaked spectrum, arising from the interplay between extended aged emission at low frequencies and compact energetic components at high frequencies.}
\label{fig:FullRadioSpec}
\end{figure}

While spectral indices provide a convenient first-order diagnostic, the full shape of the radio spectrum contains richer information about the physical conditions in the jet and the possible evolutionary stage of a blazar. As emphasized by \citet{Kerrison2024}, blazar radio spectra can exhibit diverse morphologies driven by synchrotron self-absorption, radiative cooling, or episodic core reactivation. To capture this diversity, we classified the spectra of our sample into four main morphological types, illustrated in Figure~\ref{fig:FullRadioSpec} with representative observational data.

The simplest morphology is a power law (nearly linear in log--log space), consistent with optically thin synchrotron emission \citep{Urry1995, ODea2021}. This shape indicates a stable distribution of relativistic electrons within the jet, with negligible absorption or secondary components. Over time, radiative cooling of high-energy electrons shifts the emission towards lower frequencies while preserving the overall slope, producing a canonical power-law spectrum. Such spectra are often associated with blazars in relatively steady activity phases.

A second morphology exhibits a convex, peaked shape, with flux density rising at low frequencies, reaching a maximum, and then declining at higher frequencies \citep{Snellen1998}. This curvature is usually attributed to synchrotron self-absorption: the plasma is opaque at low frequencies but becomes transparent as the frequency increases, revealing the standard synchrotron slope \citep{Risaliti2002, ODea2021}. Peaked spectra are commonly found in young or compact sources, where recent particle injection and small physical sizes enhance absorption \citep{Orienti2007, Orienti2010}.

A more complex morphology involves multiple curvatures, producing composite spectra that cannot be described by a single synchrotron component. These multicomponent spectra are interpreted as the superposition of several particle-injection episodes in the core (e.g. \citealt{Baum1990, Edwards2004, Hancock2010}). An older, extended population often dominates at low frequencies, while younger, compact components emerge at intermediate or high frequencies. This morphology, sometimes termed retriggered-peaked, indicates that the AGN has undergone recurrent activity cycles, leaving imprints of its energetic history in the radio spectrum \citep{Hogan2015, Callingham2017}.

A fourth morphology displays an inverted-peaked shape, where the flux decreases with increasing frequency, like a typical power-law, but at high frequencies, the flux starts rising again (e.g. \citealt{Hogan2015, Ballieux2024}). This unusual structure can be explained as the combination of a diffuse, aged component dominating the low-frequency emission and a compact, energetic population producing the high-frequency rise, with the intermediate depression arising from opacity effects and the interplay of the two \citep{Bicknell2018}. Such spectra may signal simultaneous contributions from large-scale relic emission and either a reactivated core or compact components recently ejected from it \citep{Brienza2020}.

In the literature, previous works have determined the shape of the radio SED using different frequency coverages. Specifically, some studies have used between 1 and 1.5 decades in frequency (e.g. \citealt{Kovalev2002}; \citealt{Kukreti2023}; \citealt{Shuvo2024}; \citealt{Kukreti2024}), while others have used 2 or more decades (e.g. \citealt{Dey2022}; \citealt{Kerrison2024}). To be conservative, we restricted our classification of spectral shapes to sources with a frequency coverage of at least 2 decades, excluding cases that could not be classified due to complex or irregular shapes, possibly caused by the superposition of multiple emission regions within the jet, underscoring the wide spectral diversity present in blazars. Following this criterion, we were able to classify 74\% of the sample: 23\% exhibit power-law spectra, 13\% show peaked spectra, 60\% display retriggered-peaked morphologies, and 4\% fall into the inverted-peaked category. A composite classification of individual sources is available in the Zenodo dataset \citep{GuerreroGonzalez2026}.


\section{RESULTS AND CONCLUSIONS}
\label{sec:Results and Conclusions}

\begin{figure}[ht!]
\centering
\includegraphics[width=\columnwidth]{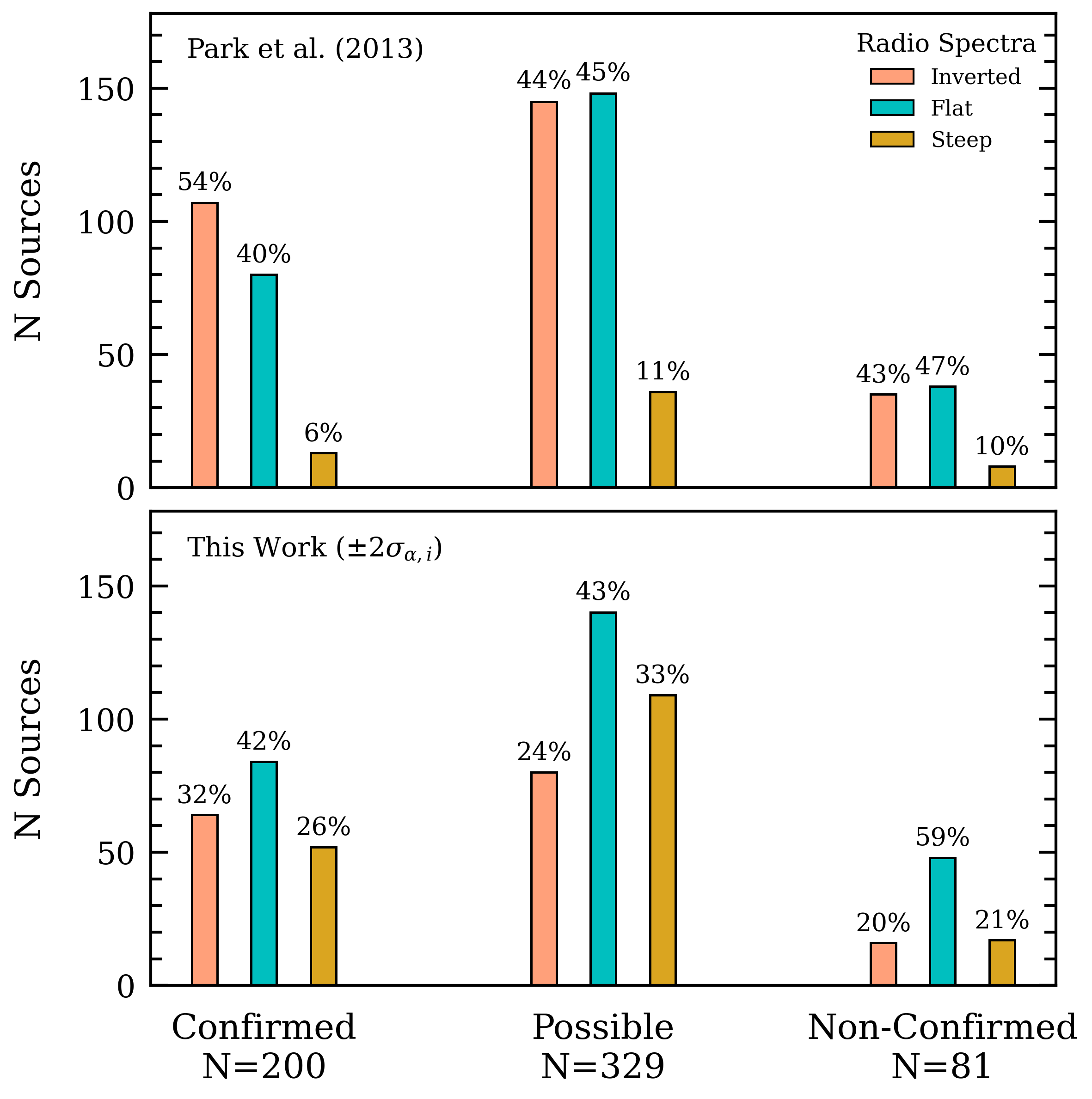} 
\caption{Comparison of radio spectral classifications under different schemes. Top panel: distributions obtained using the Park et al. (2013) criterion. Bottom panel: distributions derived from the source-by-source $2\sigma_{\alpha,i}$ classification adopted in this work. In both cases, results are shown separately for the three populations defined in this work: Confirmed BZQ, Possible BZQ, and Non-Confirmed BZQ.}
\label{fig:FinalResults}
\end{figure}

Figure~\ref{fig:FinalResults} summarizes the distribution of radio spectral classifications across the three source categories defined in this work: Confirmed BZQ, Possible BZQ, and Non-Confirmed BZQ. The top panel presents results obtained using the classical scheme of \citet{Park2013}, while the bottom panel shows the distributions derived from our revised classification.

Under the \citet{Park2013} framework, a surprisingly large fraction of sources in all three populations are assigned to the inverted category, in many cases exceeding the fraction of Flat-spectrum sources. For example, more than half of the Confirmed BZQs are labeled as inverted, while the flat category remains around 40\%. A similar trend is observed in the Possible and Non-Confirmed BZQ groups, where inverted and flat classifications appear in nearly equal proportions. These results expose a key limitation of the classical thresholds: many spectra with $\alpha$ values very close to zero are artificially assigned to the inverted regime, even though their shapes are virtually indistinguishable from flat spectra.

In contrast, under the adopted source-by-source $2\sigma_{\alpha,i}$ criterion, flat spectra become the dominant class in all three populations. This is expected for a sample of quasar-type blazars and confirms that many sources classified as inverted by the nominal sign of $\alpha$ are in fact consistent with $\alpha=0$ within their individual uncertainties. Nevertheless, a substantial fraction of sources still shows statistically meaningful departures from strict flatness, appearing as either steep or inverted. This result indicates that, even after incorporating individual measurement uncertainties, the radio spectra of BZQs are not uniformly flat.

\begin{figure}[ht!]
\centering
\includegraphics[width=\columnwidth]{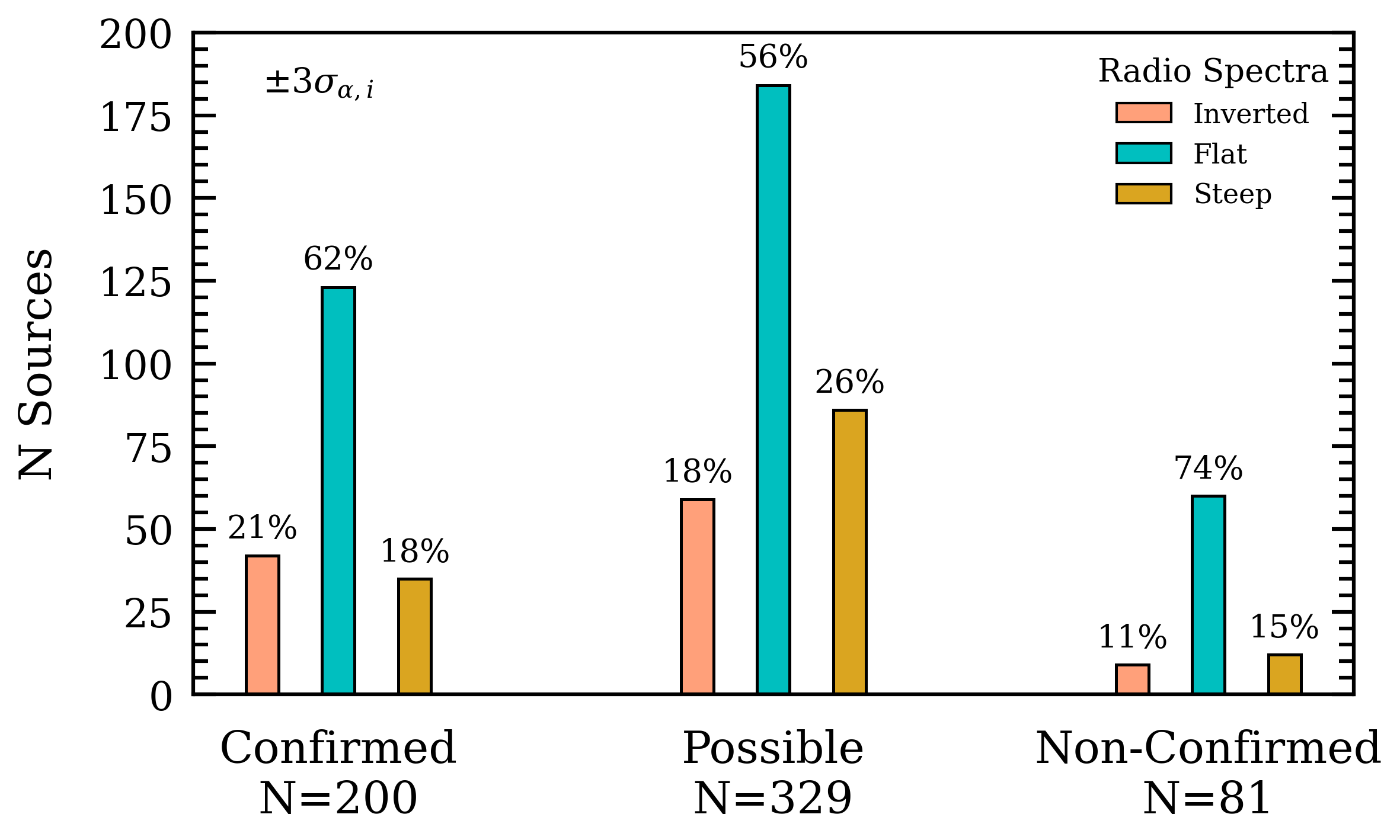} 
\caption{Radio spectral classifications obtained using an alternative source-by-source $3\sigma_{\alpha,i}$ threshold. This figure is shown for transparency and comparison with the adopted $2\sigma_{\alpha,i}$ classification shown in Fig.~\ref{fig:FinalResults}. As expected, the more conservative $3\sigma_{\alpha,i}$ criterion increases the fraction of sources classified as flat, but the qualitative population-level trends remain consistent with the adopted scheme.}
\label{fig:three_sigma_classification}
\end{figure}

For transparency, Fig.~\ref{fig:three_sigma_classification} shows the corresponding distributions obtained using an alternative source-by-source $3\sigma_{\alpha,i}$ threshold. As expected, the more conservative $3\sigma_{\alpha,i}$ criterion increases the fraction of flat spectra in all three populations; nevertheless, the resulting classification changes affect only 15.6\% of the sample. However, the qualitative result remains unchanged: the BZQ population exhibits a broad range of radio spectral behavior, and a non-negligible fraction of sources still departs from strict flatness. The adopted $2\sigma_{\alpha,i}$ threshold therefore provides a balanced criterion for the phenomenological goals of this study, avoiding both the overclassification of near-zero spectra as inverted and the excessive retention of statistically meaningful non-flat spectra in the flat category.

This comparison underscores the strength of the revised scheme, which accounts for measurement uncertainties and produces classifications that align better with spectral shapes. Nevertheless, not all BZQs exhibit flat spectra. Although flat spectra remain the majority, a non-negligible fraction displays steep or inverted morphologies, directly challenging the long-standing assumption that all \textit{Flat-spectrum radio quasars} necessarily possess flat spectra.

Moreover, it has been shown in individual sources that the radio spectral index evolves with time, as expected in AGNs due to intrinsic jet variability (e.g., 3C 279: \citealt{Patino2019}; 3C 454.3: \citealt{Amaya2021}; B2 1633+382: \citealt{Amaya2022}; PKS 1510-089: \citealt{Amador2024}). In some cases, these changes can be remarkably large, with the spectral index transitioning from positive (inverted) to negative (steep) values and back again over different epochs. This behavior indicates that classifications such as flat, inverted, or steep are not necessarily stable properties of a source, but can depend on the observing epoch.

On this basis, we conclude that the term \textit{Flat-spectrum radio quasar} is potentially misleading and should not be applied indiscriminately. A more accurate designation is simply BZQ, which recognizes the quasar nature of these blazars without enforcing a spectral property that is not universally satisfied. This terminology better reflects the observed spectral diversity and prevents the implicit assumption that all such sources exhibit flat radio spectra. The BZQ designation is already in use by multiple groups (e.g. \citealt{Xiong2015}; \citealt{Chen2015}; \citealt{Chen2016}), particularly by the \textit{Fermi} Collaboration.

While our revised classification provides a more robust framework, several open questions remain. Extending the analysis to broader frequency ranges, from MHz to tens of GHz, will reveal whether spectral diversity persists across different observing windows. Incorporating time-domain information is also essential to evaluate whether sources transition between spectral regimes and to test the stability of classifications across epochs. In parallel, future work should explore the physical drivers such as jet orientation, Doppler boosting, synchrotron self-absorption, or environmental effects that produce steep or inverted spectra among BZQs. Since we have seen that the jet orientation at pc scales does not necessarily align with the jet orientation at kpc scales, using the observed jet orientation as the sole classification criterion might result in multiple blazars (with pc-scale jets aligned with our line of sight) not being classified as such due to the kpc-scale jet orientation. Therefore, a unified approach that combines radio spectral morphology with optical and $\gamma$-ray properties could lead to a more physically motivated taxonomy of blazars. Finally, applying this scheme to the full 5BZCAT catalog and to next-generation surveys (e.g., SKA; \citealt{Dewdney2009, Braun2019}; ngVLA; \citealt{Murphy2018, Selina2018a, Selina2018b}) will provide statistically robust insights into the true distribution of spectral classes. Together, these efforts refine our understanding of blazar populations and contribute to a more consistent and physically motivated classification framework.

In addition to the analysis presented here, this work provides a publicly available and homogeneous database of broadband radio spectra for a large sample of quasar-type blazars. The full set of radio spectra and derived classifications is intended to serve as a reference dataset for future studies focused on global radio spectral behavior, specific spectral classes, or population-based analyses within the rest-frame frequency range defined in this work. We anticipate that this resource will be useful for a broad range of investigations on blazar radio emission beyond the scope of the present study.

\section{ACKNOWLEDGEMENTS}

We thank the anonymous referee for the constructive comments that helped to improve the manuscript. We thank Andrei Lobanov and Sergio Dzib for valuable discussions and assistance with the interpretation of radio spectra. J.U.G.-G. gratefully acknowledges support from \textit{La Secretaría de Ciencia, Humanidades, Tecnología e Innovación} (SECIHTI) Ph.D. fellowship program. This research was carried out within the framework of the Max Planck Institute for Radio Astronomy (MPIfR)–Mexico Max Planck Partner Group, directed by V.M.P.-A. Funding for the Sloan Digital Sky Survey IV has been provided by the Alfred P. Sloan Foundation, the U.S. Department of Energy Office of Science, and the Participating Institutions. SDSS-IV acknowledges support and resources from the Center for High Performance Computing at the University of Utah. The SDSS website is \url{http://www.sdss4.org}. SDSS-IV is managed by the Astrophysical Research Consortium for the Participating Institutions of the SDSS Collaboration including the Brazilian Participation Group, the Carnegie Institution for Science, Carnegie Mellon University, Center for Astrophysics $|$ Harvard \& Smithsonian, the Chilean Participation Group, the French Participation Group, Instituto de Astrofísica de Canarias, The Johns Hopkins University, Kavli Institute for the Physics and Mathematics of the Universe (IPMU) / University of Tokyo, the Korean Participation Group, Lawrence Berkeley National Laboratory, Leibniz Institut für Astrophysik Potsdam (AIP), Max-Planck-Institut für Astronomie (MPIA Heidelberg), Max-Planck-Institut für Astrophysik (MPA Garching), Max-Planck-Institut für Extraterrestrische Physik (MPE), National Astronomical Observatories of China, New Mexico State University, New York University, University of Notre Dame, Observatário Nacional / MCTI, The Ohio State University, Pennsylvania State University, Shanghai Astronomical Observatory, United Kingdom Participation Group, Universidad Nacional Autónoma de México, University of Arizona, University of Colorado Boulder, University of Oxford, University of Portsmouth, University of Utah, University of Virginia, University of Washington, University of Wisconsin, Vanderbilt University, and Yale University. The ZTF forced-photometry service was funded under the Heising-Simons Foundation grant \#12540303 (PI: Graham). This paper is based (in part) on results obtained with LOFAR-ERIC equipment. LOFAR \citep{vanHaarlem2013} is the Low Frequency Array designed and constructed by ASTRON. The FIRST project has been supported by grants from the National Aeronautics and Space Administration, the National Science Foundation, NATO, the National Geographic Society, the Sloan Foundation, the Institute of Geophysics and Planetary Physics, Columbia University, and Sun Microsystems.This research has made use of data from the Astrogeo VLBI FITS image database maintained by L. Petrov, as well as the SED Builder tool developed at the ASI Space Science Data Center (SSDC). We also acknowledge the use of the CATS database, maintained by the Special Astrophysical Observatory of the Russian Academy of Sciences (SAO RAS). We acknowledge the use of MagicPlot software (version 3.0.1; \url{http://www.magicplot.com}) for data analysis.


\renewcommand{\refname}{REFERENCES}
\bibliography{References}


\end{document}